# Internet of Things (IoT) Application Model for Smart Farming

Jagruti Sahoo
*Dept. of Computer Science and Mathematics*
South Carolina State University
Orangeburg, USA
jsahoo@scsu.edu

Kristin Barrett
*Dept. of Computer Science and Mathematics*
South Carolina State University
Orangeburg, USA
kbarret1@scsu.edu

*Abstract*—Smart Farming has brought a major transformation in the agriculture process by using the Internet of Things (IoT) devices, emerging technologies such as cloud computing, fog computing, and data analytics. It allows farmers to have real-time awareness of the farm and help them make smart and informed decisions. In this paper, we propose a distributed data flow (DDF) based model for the smart farming application that is composed of interdependent modules. We evaluate the proposed application model using two deployment strategies: cloud-based, and fog-based where the application modules are deployed on the fog and the cloud data center respectively. We compare the cloud-based and fog-based strategy in terms of end-to-end latency and network usage.

## I. Introduction

Internet of Things (IoT) is a revolutionary technology that allows users to interact with smart and connected devices in novel ways. Smart farming has recently emerged as a popular IoT domain that involves data-driven and automated farm processes using a variety of IoT sensors from soil sensors, temperature/humidity sensors, pH probe to actuators such as robots and drones [1][2][3]. With increased agricultural yield, reduced cost, and optimal resource usage, smart farming can solve the significant issue of meeting the global food demand by 2050 [1]. Cloud and fog computing paradigms have been leveraged to satisfy the computational need of IoT applications. While cloud supports resource-intensive applications, fog nodes can run delay-sensitive applications by bringing virtualized resources to the network edge [4].

Distributed Data Flow (DDF) is an effective approach to model IoT applications as a collection of interdependent modules that can be deployed in a distributed computing environment [5][6]. However, the efforts have been mostly limited to smart city [7], and smart healthcare [8] domains. Since smart farming involves different functionalities, it is necessary to investigate the DDF approach for modelling smart farming applications. In this paper, we propose a DDF based model for soil management application that consists of a number of modules with varying resource needs. Moreover, the proposed model has multiple application loops, each of which generates a specific type of result (e.g., status, alerts, and recommendation) for the end-user. We evaluate the performance of the proposed model considering both cloud-based and fog-based deployment.

The remainder of this paper is organized as follows. Section 2 presents the IoT application model. Section 3 discusses the experimental results. We draw conclusions and outline some future works in Section 4.

## II. Proposed IoT Application Model

We design an IoT application to provide soil management functions using the distributed data flow (DDF) approach [5]. The DDF model of the application is represented as a directed acyclic graph (DAG) shown in Fig. 1. The modules are connected through a directed edge that shows the data flow, also known as tuple. The IoT application receives data from the soil sensors and send the processed data to the end-user device denoted as Display in Fig. 1. The application consists of five modules: 1) *Sense*: This module collects data (tuple $t_1$) from sensors, removes incomplete and noisy information, and sends the filtered data (tuple $t_2$) to the data aggregation module, 2) *Data Aggregation:* This module combines the filtered data by using temporal/spatial aggregation to remove redundancy. It sends tuples $t_3$, $t_4$, and $t_5$ to status generation, event detection, and the soil analytics modules, respectively, 3) *Status Generation:* It provides users with real-time status of soil being monitored in the desired format, 4) *Event Detection:* This module processes the tuple $t_4$ and sends critical alerts (tuple $t_7$) to the end-user on detecting abnormal soil condition such as low moisture, 5) *Soil Analytics:* This is a lightweight analytic module that provides recommendations for optimal usage of water.

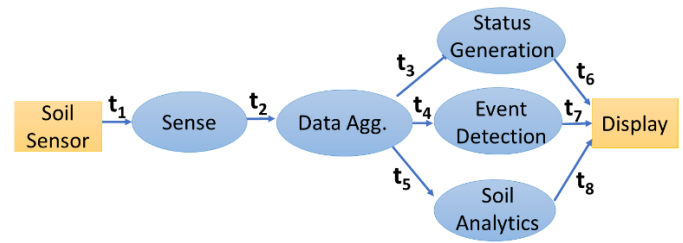

Fig. 1 DAG of Soil Management Application

## III. Experimental Results

We evaluate the proposed IoT application model using iFogSim tool [8]. We consider a network topology shown in Fig. 2 that consists of a cloud server and a hierarchical fog [9]. The Tier-2 fog server has more capacity than the Tier-1 fog servers. Sensors and the end-user device communicate directly with the tier-1 fog servers. We increase the number of sensors per tier-1 fog server from 3 to 21. Table I shows the CPU capacity of the cloud and fog servers. Table II shows the CPU requirement of application modules. We evaluate the application model using two deployment strategies: 1) cloud-based, and 2) fog-based. In cloud-based deployment, all modules are placed on the cloud data center. In Fog-based deployment, we place the 'Sense' and "Data Aggregation" module on tier-1 fog servers. The modules "Status Generation", "Event Detection", and "Soil Analytics" are placed on the tier-2 fog server. We consider two performance metrics: 1) End-to-end latency and 2) network usage. End-to-end Latency is the time elapsed from the instant the sensor generates a tuple until it is consumed by the end-user device. It includes network delay and execution delay. Network usage represents the number of data bytes transmitted through the network during the execution of the application.

Fig. 3 shows that network usage increases with respect to the number of sensors in both strategies. As more tuples are generated, they bring more load to the network. We observe that network usage of cloud-based strategy is much higher than that of fog-based strategy irrespective of the number of sensors.

Fig. 4 shows the impact of the number of sensors on end-to-end latency. We observe that the latency of the cloud-based strategy is significantly higher than that of fog-based strategy. Also, each strategy experiences a slight increase in the end-to-end latency with an increase in the number of sensors. This is due to the increase in waiting time before the tuples are processed by the respective modules, resulting in higher latency.

## IV. CONCLUSION AND FUTURE WORK

In this paper, we propose a model for IoT based soil management application and evaluate the model using cloud-based and fog-based deployments. Our experimental results show that fog-based strategy outperforms the cloud-based strategy both in terms of latency and network usage. In the future, we will perform extensive evaluation by considering concurrent execution of the multiple smart farming applications including soil management, crop management, and farm surveillance. Although fog servers offer lower delay and save bandwidth, application modules with higher resource requirements need to be placed on the cloud data center. We will design an algorithm to place the modules of concurrent applications on cloud/fog server in a way that ensures efficient resource utilization and satisfy the quality of service (e.g., latency) requirement of the applications.

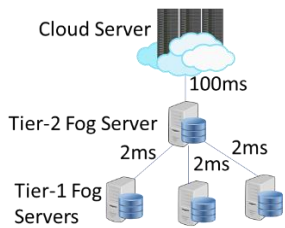

Fig. 2 Network Topology (Latency is shown for each link)

Table I CPU Capacity

| Node | CPU Capacity (MIPS) |
|---|---|
| Cloud Server | 40960 |
| Tier-2 Fog Server | 8192 |
| Tier-1 Fog server | 6144 |

Table II CPU Requirement

| Module | CPU Requirement (MIPS) |
|---|---|
| Sense | 500 |
| Data Aggregation | 600 |
| Status Generation | 500 |
| Event Detection | 1200 |
| Soil Analytics | 1200 |

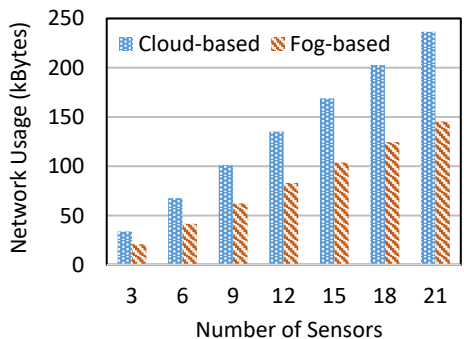

Fig. 3. Network Usage

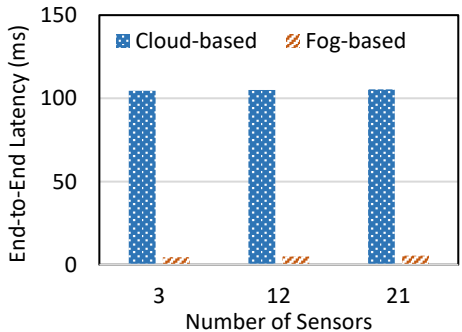

Fig. 4. End-to-End Latency

ACKNOWLEDGMENT

This work is supported by the National Institute of Food and Agriculture, United States Department of Agriculture, Evans-Allen project number SCX-314-02-19 .